\documentstyle[preprint,revtex]{aps}
\begin{document}
\draft
\begin{title}
Coherent `ab' and `c' transport theory of high-$T_{c}$ cuprates
\end{title}
\author
{A.S. Alexandrov$^{(a,b)}$, V.V. Kabanov$^{(b,c)}$  and N.F. Mott$^{(b)}$}
\begin{instit}
{$^{(a)}$Department of Physics, Loughborough University, Loughborough
LE11 3TU, U.K.

$^{(b)}$ IRC in Superconductivity, University of Cambridge, Cambridge CB3
OHE, U.K.

$^{(c)}$ Frank Laboratory of Neutron Physics, JINR, Dubna, Russia}
\end{instit}

\begin{abstract}
We propose a microscopic theory of the  `$c$'-axis and in-plane
transport of copper oxides based on the bipolaron theory and the
Boltzmann
kinetics.  The fundamental relationship
between the  anisotropy
 and the spin susceptibility is derived,
 $\rho_{c}(T,x)/\rho_{ab}(T,x)\sim x/\sqrt{T}\chi_{s}(T,x)$.
The temperature $(T)$ and doping $(x)$ dependence of the in-plane,
$\rho_{ab}$ and
out-of-plane, $\rho_{c}$  resistivity and the spin susceptibility,
$\chi_{s}$ are found in a remarkable
agreement with the experimental data  in underdoped, optimally and
overdoped $La_{2-x}Sr_{x}CuO_{4}$ for the entire temperature regime from
$T_{c}$ up to
$800K$. The normal state gap is explained and its doping and temperature
dependence is clarified.
\end{abstract}

\narrowtext

The absolute value and qualitatively different temperature dependence of
the in-plane and $c$ -axis resistivity \cite{coo} as well as the normal
state gap observed with NMR, neutron scattering, thermodynamic and
kinetic measurements in copper based high-$T_{c}$ oxides are recognised
now as the key  to our understanding of the high-$T_{c}$ phenomenon
\cite{and,mot}. By the use of the room-temperature values of the $ab$ and
$c$ -axis conductivities of the prototypical cuprate $La_{2-x}Sr_{x}CuO_{4}$
at the optimal doping,
  $\sigma_{ab}\simeq 2\times
10^{5}\Omega^{-1}m^{-1}$ and $\sigma_{c}\simeq 10^{3}\Omega^{-1}m^{-1}$,
$x=0.15$,\cite{nak}
one obtains
\begin{equation}
{E_{F}\tau\over{\hbar}}={\pi\hbar d \sigma_{ab}\over{e^{2}}}\simeq 1
\end{equation}
and the ratio of the mean-free path $l_{c}$ to the inter-plane distance
$d$  as
\begin{equation}
{l_{c}\over{d}}={2\pi \hbar \sigma_{c}\over{e^{2}k_{F}^{2}d}}< 0.01
\end{equation}
where $E_{F}$ and $\hbar k_{F}$  are the Fermi energy and $2D$ Fermi
momentum, respectively, and $\tau$ is the transport relaxation time.
 This estimate as well as the semiconductor-like behavior of
$\rho_{c}(T)$
contrasting with the linear $\rho_{ab}(T)$ do not agree with any
Fermi-liquid description
\cite{and}. Another challenging problem is a
three-dimensional coherent superconducting state of these quasi-two
dimensional
conductors which is hardly compatible with several phenomenological
models \cite{coo,pin}, based on the assumption that the
c-axis transport is  $incoherent$. To meet this challenge
some authors\cite{and} alleged
the spin-charge separation  abandoning the Fermi-liquid and Boltzmann
approach.

At the same time the results of the  kinetic \cite{buc,ito,car,mag} and
the heat
capacity\cite{lor} measurements led us\cite{alemot} to the conclusion
that the
so-called spin-gap observed previously in the magnetic susceptibility
\cite{tak,ros,tra}
(in the spin channel) belongs in fact to the charge carriers,
which are inter-site  bipolarons. In
particular, the quantitative explanations of the temperature dependence
of the
NMR line-width and the
linear in-plane resistivity\cite{ale2} as well as of the Hall effect
\cite{alebra} were
proposed. Several authors \cite{ito,nak,pin}
attributed the `semiconducting' temperature dependence $\rho_{c}(T)$ to
the
`normal state gap' in the density of states.
The comprehensive analysis by Batlogg and co-workers
\cite{bat} revealed transport features incompatible with
the spin-charge separation and  the doping dependence of the
normal state gap as well the semiconducting-like doping dependence of
resistivity. It became clear that high-$T_{c}$ oxides are doped
semiconductors rather than metals, irrespective to the level of
doping\cite{ale}.

In this letter the $coherent$  theory of $
\rho_{c}(T,x), \chi_{s}(T,x)$ and of the normal state gap, $\Delta(T,x)$
is developed based on the
bipolaron theory of high-temperature superconductivity by
Alexandrov and Mott  \cite{alemot}.

   We expect that small polarons are
    paired  $above$ $T_{c}$
    in strongly correlated  Mott-Hubbard insulators with the
   electron-phonon coupling constant above an intermediate value
    $\lambda\geq 0.5$ \cite{alemot2}.
   The ground and low-energy states are well described by the
   mixture of the intersite in-plane singlet  pairs (small bipolarons)
and
   thermally excited polarons.
   Above $T_{c}$,
   which is the condensation temperature of the charged Bose-gas,
  all carriers are nondegenerate.  Singlets
   tunnel along the plane with an effective mass $m^{**}$ of the order
   of a single-polaron mass $m^{*}$ as shown by one of us \cite{ale}.
   However, their $c$-axis tunneling can be only
   Josephson -like involving simultaneous hopping of two holes.
   Therefore the singlet $c$ -axis mass is strongly enhanced,
   $m^{**}_{c}>>m^{**}\sim m^{*}$. In that way we explain a
   large transport
   anisotropy of copper oxides at low temperatures when  polarons are
   frozen out which is difficult to understand from the band-structure
   calculations alone. The crucial point of our theory is that  polarons
dominate in the $c$-axis
   transport  at intermediate and high temperatures because they
   are much lighter in the $c$-direction than bipolarons (see below).
 Along the planes they
   propagate with about the same
   effective mass  as  singlets. Therefore their contribution  to the
$ab$
   transport is small at any temperature due to their low density
   compared with the density of  bipolarons.   As a
   result we have a mixture of the $nondegenerate$ quasi-two dimensional
   spinless bosons and the thermally excited fermions,
   which are capable of propagating along the  $c$ -axis. Because only
polarons
   contribute to the spin
   susceptibility there is a  fundamental relationship
   between the anisotropy and the magnetic susceptibility.

    Applying
   the Boltzmann theory we obtain the following kinetic coefficients
   ($\hbar=k_{B}=1$)
\begin{equation}
\sigma_{ab}(T,x)=-\int_{0}^{\infty} dE \sigma_{b}(E) {\partial
f_{b}\over{\partial E}} ,
\end{equation}
\begin{equation}
\sigma_{c}(T,x)=-\int_{0}^{\infty} dE \sigma_{pc}(E) {\partial
f_{p}\over{\partial E}} ,
\end{equation}
and
\begin{equation}
\chi_{s}(T,x)=-2\mu_{B}^{2}\int_{0}^{\infty} dE N_{p}(E) {\partial
f_{p}\over{\partial E}}.
\end{equation}
where $f_{b}=[y^{-1}exp(E/T)-1]^{-1}$ and
$f_{p}=[y^{-1/2}exp(E/T+\Delta/2T)+1]^{-1}$
are the
bipolaron and polaron distribution functions, respectively, with
$y=exp[\mu(T,x)/T]$, $\mu(T,x)$ the chemical potential,
 and $\mu_{B}$ the Bohr magneton. The bipolaron binding energy
 $\Delta$ is expected to be of the order of a few hundred $K$
 \cite{alemot}. Therefore polarons are not degenerate at any
temperatures.
 Above $T_{c}$ bipolarons are also not degenerate, so that
 \begin{equation}
 f_{b}\simeq y exp\left(-{E\over{T}}\right),
 \end{equation}
 and
 \begin{equation}
 f_{p}\simeq \sqrt{y}exp\left(-{E+\Delta/2\over{T}}\right).
 \end{equation}
 If the scattering mechanism is the same for polarons and bipolarons the
 ratio of the differential conductivities is independent of the energy
 and doping
 \begin{equation}
 {\sigma_{b}(E)\over{\sigma_{pc}(E)}} \equiv A=constant.
 \end{equation}
 There is a large difference in the values of the $pp\sigma$ and $pp\pi$
 hopping integrals between different oxygen sites. Therefore we expect
a highly anisotropic  polaron energy
 spectrum \cite{ale} with a quasi one -dimensional polaron density of
states
 as observed with the high resolution $ARPES$ \cite{gof}
 \begin{equation}
 N_{p}(E)\simeq {1\over{2\pi}} \sqrt{m^{*}\over{2E}}.
 \end{equation}
 Then  the $c$-axis resistivity as well as
 the spin susceptibility is expressed as
 \begin{equation}
 {\rho_{c}(T,x)\over{\rho_{ab}(T,x)}}=A\sqrt{y}
 exp\left({\Delta\over{2T}}\right)
\end{equation}
and
\begin{equation}
 \chi_{s}(T,x)=\mu_{B}^{2}\sqrt{ym^{*}\over{2\pi T}}
exp\left(-{\Delta\over{2T}}\right).
 \end{equation}
The chemical potential, $y=2\pi
 n_{b}(T,x)/Tm^{**}$,  is calculated
 by taking into account the Anderson localisation of bipolarons in a
 random potential. Within a `single well-single particle'
 approximation \cite{alebra} the bipolaron density $n_{b}(T,x)$ appears
to be linear in
 temperature and doping , $n_{b}(T,x)\sim n_{L}T$, with the total  number
of
  impurity levels $n_{L}$ proportional to the doping $x$.  That fits very
well the
 temperature and doping dependence of the Hall effect $R_{H}=1/2e
 n_{b}(T,x)$
 as well as the linear $ab$-resistivity in a wide
 temperature and doping  range as shown in ref.\cite{alebra}.
  As a result we find
 the temperature independent $y\sim x$ and
\begin{equation}
 {\rho_{c}(T,x)\over{\rho_{ab}(T,x)}}=constant
\times{x\over{T^{1/2}\chi_{s}(T,x)}},
\end{equation}
\begin{equation}
 \chi_{s}(T,x)=constant'\times \sqrt{x\over{T}}
 exp\left(-{\Delta(T,x)\over{2T}}\right).
 \end{equation}
 We expect a strong dependence of the binding energy on the doping
 due to
 screening, $\Delta=\Delta(T,x)$. Bipolarons are heavy nondegenerate
particles which screen
 very well the electron-phonon interaction. In fact,
  by the use of the classical expression for the inverse screening radius
  \begin{equation}
  q_{s}=\sqrt{16\pi e^{2} n_{b}(T,x)\over{\epsilon_{0}T}}
  \end{equation}
 and the static dielectric constant of $LSCO$, $\epsilon_{0}\simeq 30$
  one obtains the value of $q_{s}$ about $3 \AA^{-1}$ at room temperature
 with $n_{b}=10^{21} cm^{-3}$
  which is about three times larger than the reciprocal lattice vector
  $q_{d}$.
  The polaron (Franck-Condon) level shift $E_{p}$ is suppressed
  by the screening as $(q_{d}/q_{s})^{2}$ at large
  $q_{s}$\cite{alemot2}.
  Consequently,
   the normal state gap,
  $\Delta \simeq 2E_{p}$ , depends on
 the doping and temperature as
  \begin{equation}
  \Delta(T,x)\sim {T\over {n_{b}(T,x)}}.
  \end{equation}
 Taking $n_{b}(T,x)\sim Tx$ we arrive with the temperature independent gap,
  \begin{equation}
  \Delta={\Delta_{0}\over{x}},
  \end{equation}
 where $\Delta_{0}$ is  doping independent.
 By the use of
Eq.(12), Eq.(13) and Eq.(16) one can describe all qualitative
features of the $c$-axis resistivity and the magnetic
susceptibility  of $LSCO$   without any
fitting parameters as the comparison of Fig.1 and Fig.2 shows.  The
  linear temperature dependence of the $ab$ resistivity
  was explained within the same approach\cite{ale2,alebra}.
  Thus the $c$- axis resistivity is now understood as well. For a
  quantitative fit  one has to solve the Bethe-Salpeter equation
  for $\Delta(T,x)$ with the realistic
  interaction between polarons taking into account the polarisation of
the
  bipolaronic liquid partly localised by disorder. At temperatures close
  to the transition the Bose statistics becomes important.
  At large temperatures the finite polaron and
  bipolaron bandwidth (about $1000K$) plays some role.
  Therefore it is not surprising  that the experimental
   dependence of $\chi_{s}(T,x)$  shifts  from the
  theoretical one, Fig.1b. At the same time
  the anisotropy is quantitatively described by Eq.(12) with the
  $experimental$ values of $\chi_{s}(T,x)$, Fig.2a allowing small sample
  variations of $constant$ in Eq.(12)  within less than
  $15\%$. A  smaller  anisotropy of a heavily overdoped sample
  ($x=0.3$) in Fig.2a is due to the fact that
  polarons contribute to the $ab$ transport  when the binding energy is
  below $100 K$. Alternatively, the normal state
  gap $\Delta(T,x)$ can be determined  by the use of Eq.(11) and the
  experimental values of
  $\chi_{s}(T,x)$, Fig.2b. With the temperature independent $y(x)$ (in
  the agreement with a flat temperature
  dependence of the thermoelectric power of $LSCO$) one obtains
  \begin{equation}
  \Delta(T,x)=2T \ln {B_{\infty}\over{\sqrt{T}\chi_{s}(T,x)}},
  \end{equation}
  where $B_{\infty}=\lim \sqrt{T}\chi_{s}(T,x)$ for $T\rightarrow
  \infty$ is independent  of the doping, $B_{\infty} \simeq 5.46 \times
  10^{-6} Emu K^{1/2}/g$. The values of
   $\Delta(T,x)$,  determined
  with Eq.(17), Fig.3,  are about the same as
   Batlogg's  normal- state- gap
 temperature
  $T^{*}(x)$ \cite{bat}. Therefore we conclude that $T^{*}$ is  the
   bipolaron binding energy. A  temperature dependence of
   $\Delta(T,x)$ of the underdoped  sample  in Fig.3 is explained
   by the
   temperature dependence of the screening radius.
  From the Hall effect  \cite{bat}  one can observe
  that the bipolaron density is approximately constant below
   $200K$ for $x=0.1$ in the agreement with our previous calculations
   \cite{alebra}. Then, according to Eq.(15)  the normal state gap is
   proportional to the temperature,  $\Delta \sim T$ at low temperatures.

  The proposed kinetics of high-$T_{c}$ cuprates
   is derived from the generic Hamiltonian
  \begin{equation}
H=\sum_{<ij>}T({\bf m-n})c^{\dagger}_{i}c_{j}+\sum_{\bf q}\omega({\bf
q})(d^{\dagger}_{\bf q}d_{\bf q}+1/2)+\sum_{{\bf q},i}\omega({\bf
q})n_{i}[u_{i}({\bf q})d_{\bf q}+H.c.]+\sum_{ij}V_{ij}n_{i}n_{j},
\end{equation}
where $T({\bf m})$ is the bare hopping integral, $i=({\bf m},s)$, $j=
({\bf n},s)$ (${\bf m,n},s$ stands for sites and spin, respectively),
$n_{i}=c_{i}^{\dagger}c_{i}$, $u_{i}({\bf q})=g exp(i{\bf q}\cdot {\bf
m})/\sqrt{N}$,
$\omega({\bf q})$ are the coupling constant and the phonon frequency,
 and $V_{ij}$ is the
$direct$ (density-density) Coulomb repulsion.
 It can be
diagonalised $exactly$
 if $T({\bf m})=0$ (or $\lambda=\infty$). The ground state bipolaron
 configuration is found by the use of the lattice minimisation
 technique\cite{cat} fully taking into account the direct
 Coulomb repulsion. Then applying $1/\lambda$ perturbation
 expansion  the bipolaron effective mass tensor  is readily derived.
 In perovskite structures the in-plane
 oxygen-oxygen pairs appears to be  the ground state.
 Therefore the in-plane bipolaron
 tunneling is essentially one-particle and the in-plane effective mass
appears
to  be of the order of the small polaron mass about $10m_{e}$
 \cite{ale}. On the other hand, the $c$-axis tunneling of the bipolaron
 is only possible via
 a  Josephson-like hopping. In that case one derives
\cite{alekab,alemot2}
 \begin{equation}
 {1\over{m_{c}^{**}}}\simeq
 4t^{2}d^{2}\sqrt{2\pi\over{\omega \Delta}}exp\left[-{\Delta\over{\omega}}
 \left(1+\ln{2g^{2}\omega\over{\Delta}}\right)\right].
 \end{equation}
 Here $t=T(d)e^{-g^{2}}$ is the inter-plane polaron hopping integral.
Then  the ratio of the $c$ -axis singlet mass  to that of the
 polaron one is
 given by
 \begin{equation}
{m_{c}^{**}\over{m^{*}_{c}}}\simeq {1\over{2}}
\sqrt{\omega \Delta\over{2\pi T^{2}(d)}}exp(3g^{2})\gg 1,
\end{equation}
 which
  justifies the proposed kinetics described by  Eqs(3-5). The  isotope
effect on both $T_{c}$
  \cite{fra} and on the N\'eel temperature $T_{N}$ \cite{mor} favors the
  electron-phonon coupling as the origin of the  polaron and bipolaron
  formation in $LSCO$. The pair-distribution analysis of neutron
  scattering \cite{sen} also suggests that the `spin-gap' is consistent
  with the formation of a bipolaronic local singlet state.

  In conclusion, we have developed a coherent transport theory of
  copper based high-$T_{c}$ oxides, which  describes the
  doping and temperature dependence of the
  resistivity and the magnetic susceptibility of $LSCO$ as well as the
normal
  state gap. Our theory is selfconsistent in the sense that
   the mean-free path for bipolarons remains  larger
  than the lattice constant. The large resistivity values are due to
    a strongly enhanced effective mass and a small
  polaron density rather than to a short mean-free path. Thus the
Boltzmann kinetics
  is applied. No question arises with the three-dimensional
  superconductivity either. The Bose-Einstein condensation of
  bipolarons explains the high value of $T_{c}$ because its dependence on
  a huge $c$- axis singlet mass is only logarithmic \cite{alemot}. At
very low
  temperatures polarons are frozen out, so we expect
   the temperature independent anisotropy $\rho_{c}/\rho_{ab}$  when $T$
is low. In the
   magnetic field the normal state gap becomes smaller due to the
  spin splitting of the polaron level, so a negative $c$-axis
  magnetoresistance is expected. Both features have been recently
observed  \cite{ando}.

  We highly appreciate the enlightening discussions with Y. Ando,
  P. Edwards, J. Cooper, N.
  Hussey, W. Liang,  J. Loram, A. Mackenzie, and K. Ziebeck and the
 Royal Society financial support of one of us (VVK).

\figure{Theoretical anisotropy ($a$) and the magnetic susceptibility
($b$)
 with $\Delta_{0}=55K$.}

\figure{Experimental anisotropy  \cite{nak} ($a$)  compared with the
theoretical one by the use
of Eq.(12) with the experimental \cite{mag}  $\chi(T,x)
=\chi_{s}(T,x)+0.4\times 10^{-7} emu/g$  ($b$).}

\figure {Normal state gap as a function of temperature and
doping.}

 \end{document}